# PAPER-BASED CELL CRYOPRESERVATION

Roaa Alnemari[1], Pavithra Sukumar[1,†], Muhammedin Deliorman[1,†], and Mohammad A. Qasaimeh[1,2,*]

[1] Engineering Division, New York University Abu Dhabi (NYUAD), P.O. Box 129188, Abu Dhabi, UAE
[2] Department of Mechanical and Aerospace Engineering, New York University, New York, NY 10012, USA
[†] Equal Contribution
[*] E-mail: mohammad.qasaimeh@nyu.edu

The continuous development of simple and practical cell cryopreservation methods is of great importance to a variety of sectors, especially when considering the efficient short- and long-term storage of cells and their transportation. Although the overall success of such methods has been increased in recent years, there is still need for a unified platform that is highly suitable for efficient cryogenic storage of cells in addition to their easy-to-manage retrieval. Here, we present a paper-based cell cryopreservation method as an alternative to conventional cryopreservation methods. The method is space saving, cost-effective, simple and easy to manage, and requires no additional fine-tuning to conventional freezing and thawing procedures to yield comparable recovery of viable cells. We show that treating papers with fibronectin solution induces enhanced release of viable cells post thawing as compared to untreated paper platforms. Additionally, upon release, the remaining cells within paper lead to the formation and growth of spheroid-like structures. Moreover, we demonstrate that the developed method works with paper-based 3D cultures, where pre-formed 3D cultures can be efficiently cryopreserved.

Successful preservation of cells is extremely critical for their storage, transportation, and distribution. Conventional cryopreservation methods, such as slow freezing and vitrification, need to be precise and reliable so that cells can be used, directly or indirectly, at future times. Hence, the main goal of these methods is to put the cells, with minimal injuries, in a state where cells' metabolic activities "freeze". Among them, slow freezing has been commonly used for preserving cells for over 70 years[1, 2]. The principle of the method involves freezing the cells slowly (up to 2 hours) to sub-zero temperatures so that their metabolic activities are paused during their storage. As a result, the use of low concentration cryoprotectants and slow cooling in this method enables high cell viability and sustainable cell characteristics for various sample types, including embryos, sperms, oocytes, and primary cells and cell lines[3-5]. However, the method efficacy is considerably poor for complex multilayered samples, such as organoids, cell-scaffold constructs, and full organs due to samples' inability to completely absorb the cryoprotectants in freezing medium. On the contrary, vitrification utilizes higher concentration of cryoprotectants and expresses rapid cooling (within minutes) as a result of which cells/tissues are preserved in a "glass-like" state[1, 2, 6]. While both approaches have their merit, in general they commonly suffer from a major limitation: maintaining the broad sample stocks to ensure steady supply[7-9]. As such, to store hundreds of samples in thousands of cryogenic containers requires large-scale units. Consequently, difficulties in managing such vast amount of content contribute to loss, damage or misidentification of cells/tissues during sample retrievals[10]. For example, the American Type Culture Collection (ATCC), a leading cell line provider, nowadays maintains over 4,000 different types of cell lines. Based on a 1996 report[11], where the ATCC was estimated to manage their cell lines in about half a million cryotubes, currently we are expecting tens of millions of cryotubes are being stored around the world.

Therefore, an increased amount of efforts have been made to overcoming the long-term preservation challenges for simple to complex samples and allowing easy and affordable access to different type of cells and tissues in a timely manner[12-18]. Notably, when coupled with the ongoing continuous demand for cryopreservation of three-dimensional (3D) cell cultures in more natural settings, scaffolds with varying pore sizes have seemed to provide a good route to pursue a successful cryogenic storage of cells through the use of conventional cryopreservation methods.

3D cell cultures are demanding technology where cells are allowed to grow within scaffolds so that cell-cell/cell-environment interactions are possible; thus mimicking *in vivo* tissue microenvironment. To date, various substrates, such as glass, metals, polymers, hydrogels, and paper, have been utilized as scaffolds to match the microenvironment of the cells/tissues of interest[19]. Among them, paper has become an attractive platform in tissue engineering development, especially in 3D cell culture, offering remarkable features including biocompatibility, porosity, cost-effectiveness, and applicability for large-scale biological testing and microfabrication due to its tunable surface characteristics[20-23]. In addition, paper platforms were previously shown as supporting substrates to enhance the vitrification of mouse embryos[24] and bovine matured oocytes[25] and blastocysts[26]. For preserving tissue constructs, previous studies developed engineered porous scaffolds for cryopreservation of pre-cultured cells, such as corn starch-polycaprolactone fiber meshes for mesenchymal stromal cells[8], electrospun-polyurethane nanofiber sheets for myoblast cells[27], alginate-gelatin cryogel sponges for stem cells[9], and reticulated polyvinyl formal resins for fibroblasts[28]. The results of these studies indicated that such scaffolds offer highly protective environment for the retention of cells' viability and content during the cryopreservation. This was attributed to the biocompatibility and mechanical strength of the materials from which the scaffolds were made of, along with the porosity that facilitated the uniform distribution of cryoprotectants to the cells at different locations within scaffolds. Cumulatively, however, these scaffolds are needed to be repeatedly manufactured (i.e. engineered) for their use in relevance to cryopreservation of cells. Additionally, their applicability to cryopreserve intact (i.e. non-cultured) cells has not been reported yet. Paper, on the contrary, is ready-to-use scaffold that offers superior compatibility with the cryopreservation in addition to its remarkable potential as a 3D cell culture platform. However, it has never been utilized as a platform to cryopreserve cells.

Here, we describe a method to efficiently cryopreserve cells on paper platforms using standard slow freezing procedures, herein referred to as paper-based cryopreservation (Figure 1*a*). The method is space saving, cost-effective, simple and easy to manage, and requires no additional modification to the conventional protocols for freezing and thawing cells. In this method, cells are ubiquitous in the 3D porous environment of the paper, where paper fibers provide a natural protective and supportive environment during their preservation. As a result, after their freeze, thawed cells are efficiently released from paper with high viability rates by gently shaking the paper. In addition, preliminary results suggest that the remaining cells residing within the paper can be further utilized to create 3D cell constructs and that the method enables the cryopreservation of paper-based 3D cell culture systems. The developed method brings several advantages to the field of cryopreservation including high viability of the preserved cells (comparable to conventional slow freezing method), mechanical stability of the paper so platforms with large areas can be rolled and stored in stocks, efficiently retrieve/transport cryopreserved cells in an on-demand manner in small pieces without a need to thaw the entire platform, and a versatile 3D porous environment tailoring spheroid formations and allowing cryopreservation of pre-cultured cells in 3D. Our main motivation is to provide practical solution for the effective preservation of cells that is space saving, cost-effective, simple, and easy to manage.

## Results

**Characterization of cell release after thawing.** During the development of the paper-based cell cryopreservation, it was important to recognize that the method doesn't jeopardize the viability and functions of cells during their freeze and storage while allowing their release from papers after thaw. Therefore, our first concern was directed toward obtaining effective means to release viable cells from the paper after cryopreservation. Previously, it was reported that fibronectin, a widely used substrate in cell culturing, not only provides a biocompatible surface for the cells to retain their viability but also favors their detachment under shear[29]. With this in mind, we investigated whether fibronectin influences the release of cells from paper upon cryopreservation by comparing the remaining cells on the fibronectin-treated papers to untreated ones after thawing and shaking. To this end, paper fibers were coated with fibronectin at 10 µg/mL concentration and HeLa (a human cervical cancer cell line) cells were loaded onto papers in the order of ~$10^7$ cells/mL per $cm^2$ of paper strips (Figure 1*b*). Results suggested that fibronectin-treated papers substantially enhance viable cells' release after thawing; a ~50% increase when compared to untreated papers (Figures 1*c* & 1*d*). Next, to investigate the involvement of fibronectin-coated fibers in the release efficiency of cells and to understand if there are any biological interactions involved, we repeated the same experiments with fluorescently-labeled beads having nominal size of 20 µm. As shown in supplementary Figure S1, *a-c*, fibronectin enhanced the release of the loaded beads, while SEM images of loaded papers confirmed that beads were physically immersed within 3D environment of the paper (Figure 1*e*).

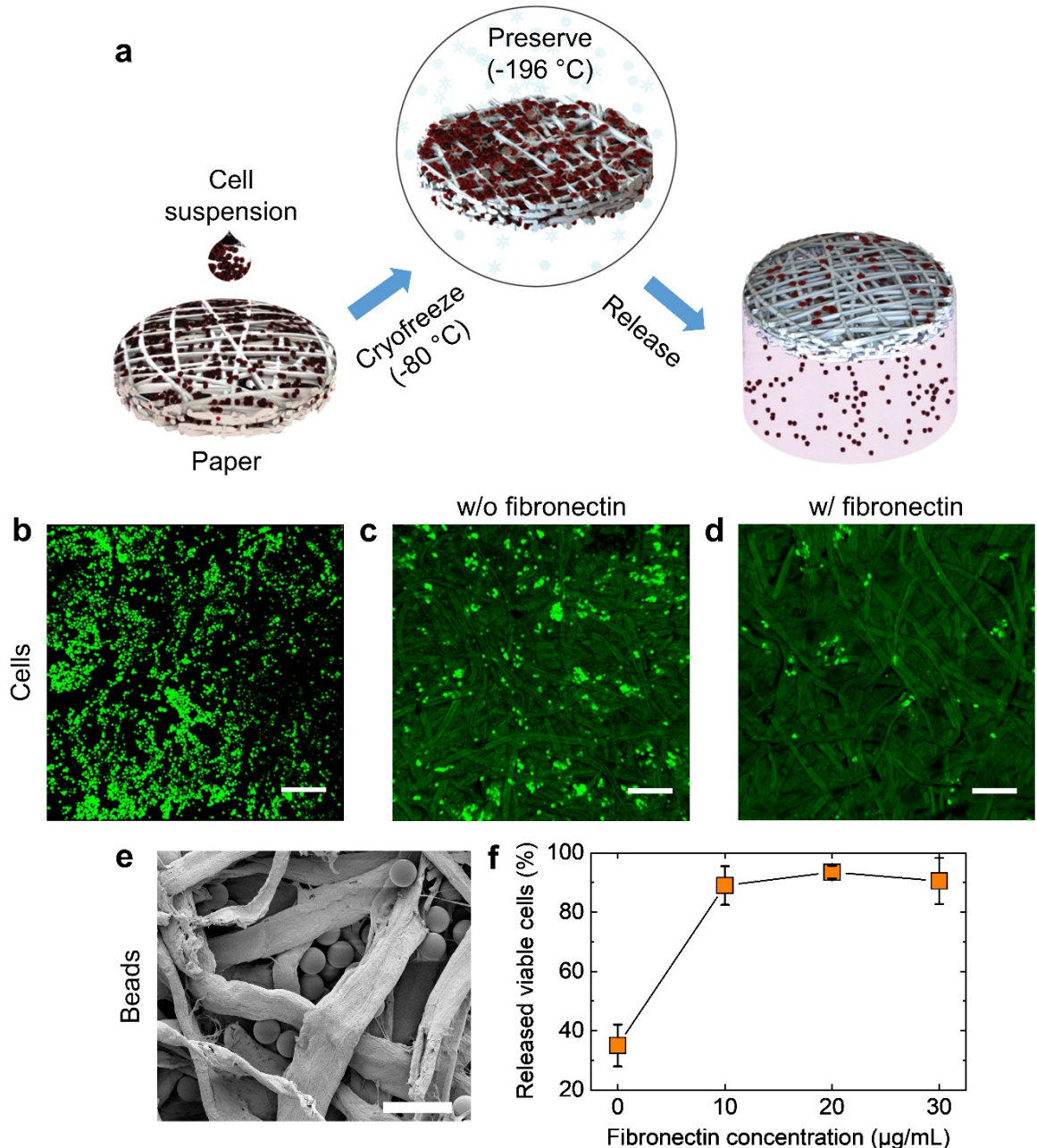

*Figure 1 | Paper-based cell cryopreservation. (a)* The methodology for utilizing paper in cryopreservation of cells is simple and robust, allowing for easy loading and efficient freezing of cells. Following their release after freeze, the remaining cells within the paper can be further grown and utilized as in vitro 3D cell constructs. *(b-d)* Z-stack confocal images show a paper with cryopreserved HeLa cells before release and the relative effect of untreated (i.e. without fibronectin) paper on the HeLa cell release to fibronectin-treated one. The scale bars are 200 µm. *(e)* SEM image of a paper shows entrapped beads, wherein mimicking the interaction of cells with the fibers and their spread in porous 3D microenvironment of the paper. The scale bar is 50 µm. *(f)* The graph represents the release efficiency of viable HeLa cells as a function of fibronectin concentration within the paper, where 10 µg/mL fibronectin solution is sufficient to release viable cells at about 90% efficiency. Values and error bars represent Mean ± S.D.

To determine whether higher fibronectin concentrations would result in more released cells, we loaded additional HeLa cells onto papers treated with 20 and 30 µg/mL fibronectin solutions at ~$10^7$ cells/mL per $cm^2$ paper strips. This, however, resulted in slightly higher released cells (58.5% and 55.5%, respectively) in comparison with their release from untreated papers (Figure 1*f*), suggesting that 10 µg/mL fibronectin concentration was sufficient for effective release of viable cells from papers following their cryopreservation steps (supplementary Figure S2).

**Cell viability and proliferation after thawing and release.** After confirming the enhanced cell release from fibronectin-coated papers and the fibronectin concentration dependency, our next concern was toward testing the compatibility of the method on maintaining cellular functions during the freeze. As opposed to cryopreservation of suspended cells in cryovials, where the distribution of cryoprotectants occurs homogeneously, cryopreservation of cells within 3D porous environment of paper is expected to be more complex since cryoprotectants need to reach cells at various locations deep in the paper[30]. To assess whether the 3D porous medium of the paper provides suitable environment to cells during their freeze, first we compared the viability of HeLa cells loaded onto papers. In terms of relative number and distribution of viable cells within papers, live/dead assays suggested that treating papers with 10 µg/mL fibronectin solution did not offer any additional advantage than untreated ones both before and after their freeze (Figure 2, *a-c*, and supplementary Figure S2, *a, b & e, f*). However, in terms of viable HeLa cell release from paper after thawing, trypan blue exclusion assays verified that fibronectin-treated papers offer superior advantage to untreated ones (Figure 2*d*). For example, the release efficiency of HeLa cells using untreated paper was about 35%, whereas the viability of released cells reached about 89% from fibronectin-treated papers.

Next, to investigate the effectiveness of the method on the release of cells other than HeLa, we conducted additional cell release experiments using PC3 (a human prostate cancer cell line), MCF-7 (a human breast cancer cell line), and JKT (a human T-cell lymphocyte cell line) cells. Regardless of the differences in their sizes and functionalities, overall results suggested that the method efficacy held for PC3, MCF-7, and JKT cells for after their release, where no significant difference on cell viability, except for cells loaded on untreated papers, was observed when compared to the conventional cryopreservation (Figure 2*d*). Similarly, as compared to the outcomes of cells on untreated papers, the release efficiency of PC3, MCF-7, and JKT cells increased by 60.5%, 28.3%, and 42.2%, respectively, when loaded onto 10 μg/mL fibronectin-treated papers (Figure 2*d*). These findings were also in agreement with after release live/dead characterization of cells remaining within papers as compared to untreated ones (supplementary Figure S2, *c, d & g, h*). Nonetheless, the release efficiencies of viable MCF-7 and JKT cells from papers were somewhat lower by 20-25% than that of recovered using conventional cryopreservation. This, in part, could be due to increased focal adhesion sites of these cells[31], and/or due to their size heterogeneity.

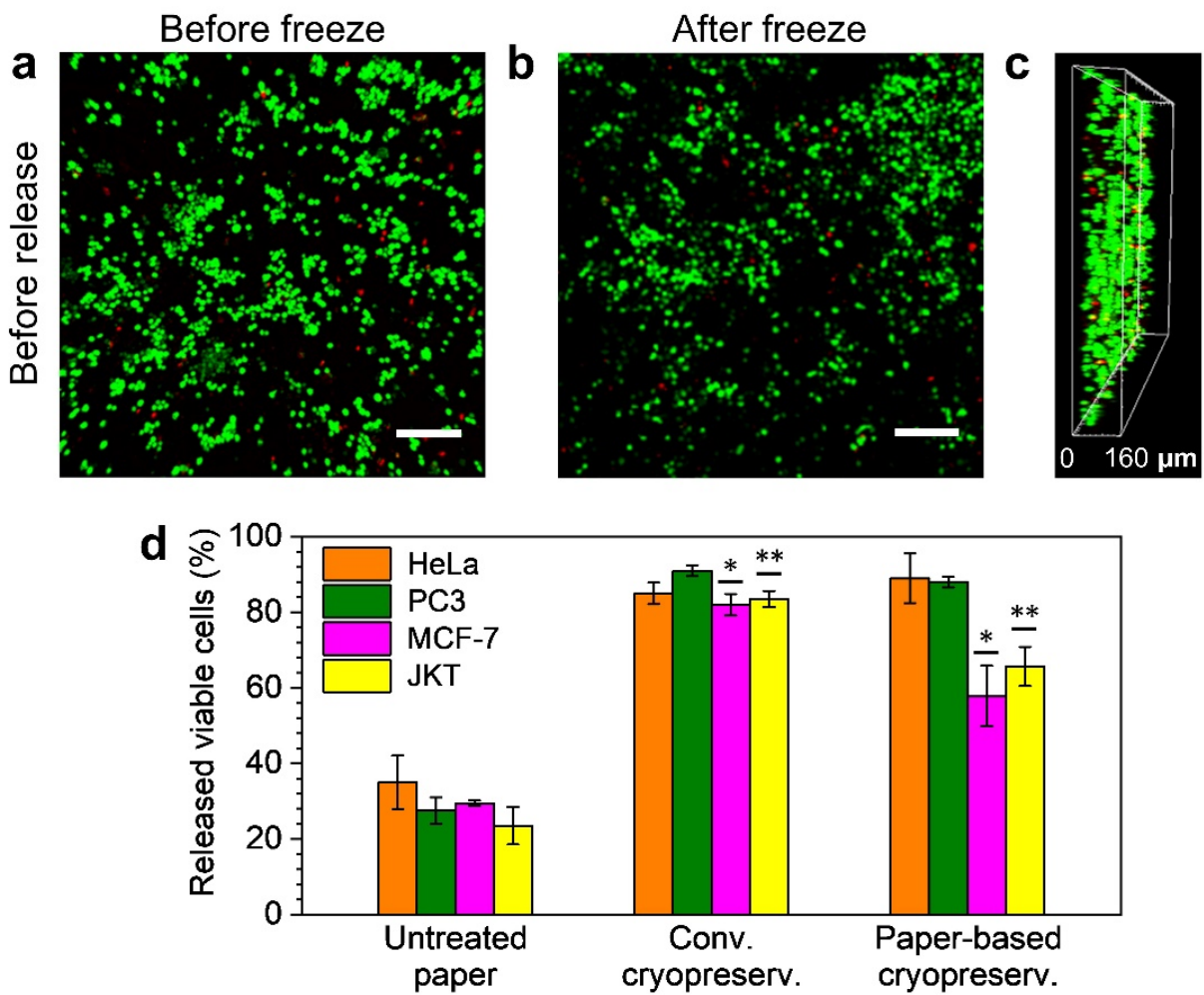

*Figure 2* | *Viability of cells cryopreserved within paper platforms. (a & b)* Z-stack confocal images of live (green) and dead (red) HeLa cells within fibronectin-treated papers confirming their unchanged relative numbers before and after freezing, respectively. Scale bars are 200 μm. *(c)* Side view of the z-stack image in *(a)* shows the depth distribution of HeLa cells within the paper. *(d)* Trypan blue exclusion assay on HeLa, PC3, MCF-7, and JKT cells reveals that paper-based cryopreservation is comparable to conventional cryopreservation in terms of recovery of viable cells following thawing. * & ** are statistically significant. Values and error bars represent Mean ± S.D.

Following live/dead assays, our desire was to investigate the cell activity after thawing and culture by proliferation analysis. Results indicated no significant differences in the cell morphologies between freshly cultured and thawed cells following their proliferation for up to 3 days. As shown in Figure 3, *a-d*, microscopic investigation of filamentous actin (F-actin) and cell nuclei staining did not show any morphological cell abnormality as compared to ones after conventional freezing, including the spreading areas of F-actin and tubulin. WST-1 cell proliferation study of HeLa cells was also performed to comparatively quantify the change in relative number of proliferated cells following thawing and release. Results confirmed that the cell culture rate of paper-based cryopreserved cells is comparable to the conventionally cryopreserved cells (Figure 3*e*).

To test if treating papers with fibronectin influences their proliferation after thawing and release, in another set of experiments we cryopreserved cells in a medium where fibronectin was used as additive. Results suggested that inclusion of fibronectin resulted in similar recovery of viable cells among cryopreservation methods (Figure 3*e*).

**Formation of 3D spheroids.** Formation of 3D tumor spheroids have become an attractive pursuit to develop and study anti-cancer drugs. As a result, various methods, such as pellet culture, liquid overlay, hanging liquid drop, and droplet-based microfluidics, have been proven to successfully facilitate easy formation of well-defined spheroids in laboratory settings[32]. In our work, however, when we allowed remaining 10-20% of cells (HeLa and MCF-7) within paper to grow in a culture medium for up to 6 days following the thaw and release, interestingly they resulted in "grape-like" spheroid growths[33, 34]. Indeed, when the surface morphology of paper fibers was investigated, the fibers exhibited wide range of sizes and large gaps are shown between the paper interlayers (Figure 4a). Hence, forming a suitable 3D structural support for aggregation preservation of cells, and as it follows, for the growth of cell clusters that resemble spheroid-like structures. Figure 4, *b & c*, shows examples of different 3D spheroid formations of the HeLa cells after their cryopreservation in paper. Following their culture for a period of 6 days, their average length grew either exponentially or linearly (Figure 4*d*) and was seemingly limited only by the pore sizes of the paper within which they resided. Such spheroid formations were easy to achieve if cells were initially loaded at high concentrations (~$10^7$ cells/mL per cm$^2$ paper strips), although at lower cell concentrations we were also able to observe clusters of cells as shown in Figure 4*e* and supplementary Figure S3. Finally, these structures were also well distributed within the paper and residing deep in the paper, where live/dead assays confirmed relatively high cell viability within them (Figure 4*f*).

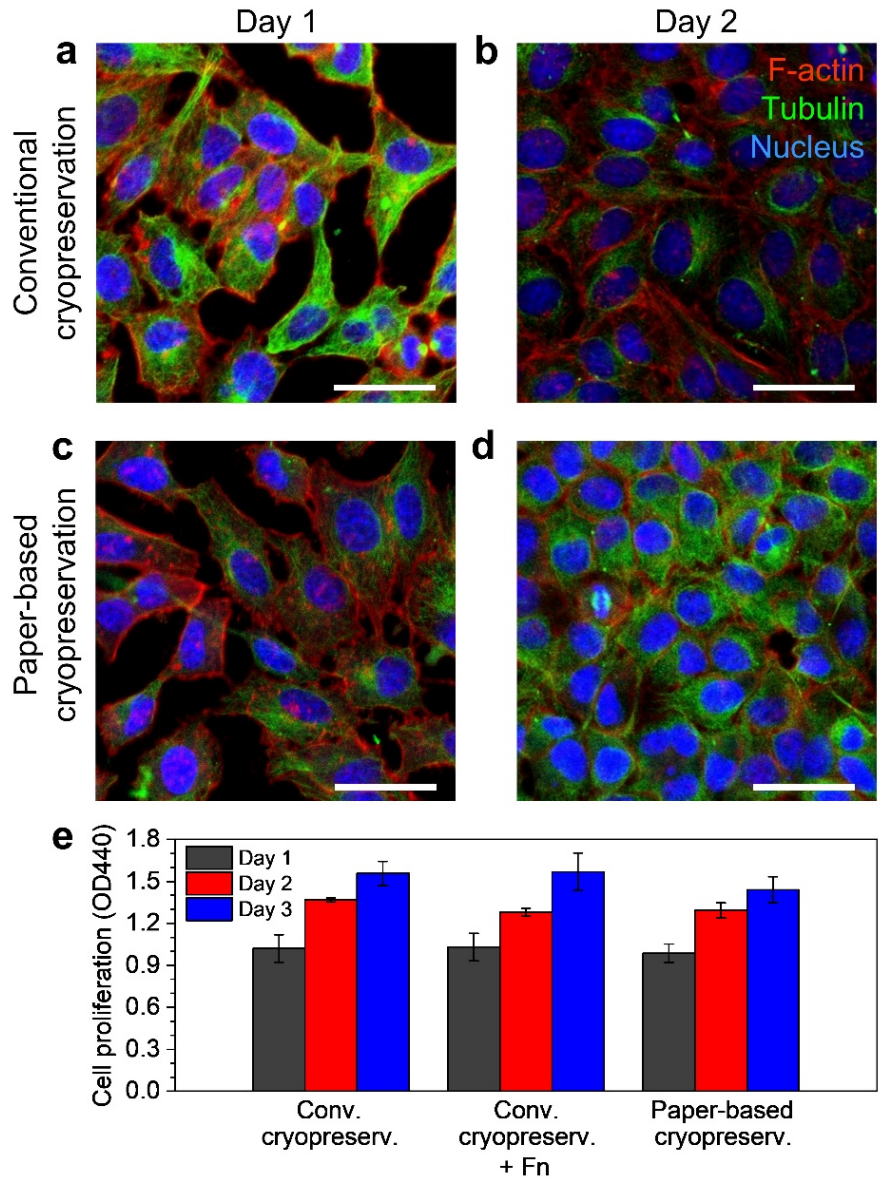

*Figure 3 | **Proliferation of cells cryopreserved within paper platforms.** (a-d) Confocal images show the Day 1 and Day 2 proliferation responses of HeLa cells following their **(a & b)** conventional and **(c & d)** paper-based cryopreservation, respectively. The spread in F-actin and tubulin confirms that paper-based cryopreservation does not hinder the morphology of cells after thawing and release. Scale bars are 20 µm. **(e)** Optical density measurements on proliferated HeLa cells for over 3 days after thawing and release reveal that paper-based cryopreservation results in comparable proliferation to conventional cryopreservation. Using fibronectin (Fn) as an additive to the cryopreservation medium did not have an influence on the change in relative numbers of proliferated cells. Values and error bars represent Mean ± S.D.*

**Cryopreservation of pre-cultured cells.** After verifying that cells loaded onto papers preserve their viability and functionality during freeze, next we explored whether paper-based cryopreservation will offer a suitable environment for the cryogenic storage of 3D cell culture systems (Figure 5*a*). For this, we investigated its functionality with cells cultured and grown within paper platforms prior to their freeze. Matrigel-suspended MCF-7 cells were loaded onto papers at concentrations of ~$10^7$ cells/mL per $cm^2$ paper strips and incubated for 3 days in culture medium (Figure 5*b*). The paper-based 3D culture was then cryopreserved for 3 days and following thawing, cells (Figure 5*c*) were additionally cultured for 6 days to fully recover[8], where their culture medium was replaced once with fresh medium every other day. Overall, our comparative proliferation images at days 4 and 9 showed that the paper-based cryopreservation maintains the integrity and viability of cells pre-cultured before cryopreservation, as confirmed by the uniformly distributed 3D cell layers growing inside the paper (Figure 5*d*).

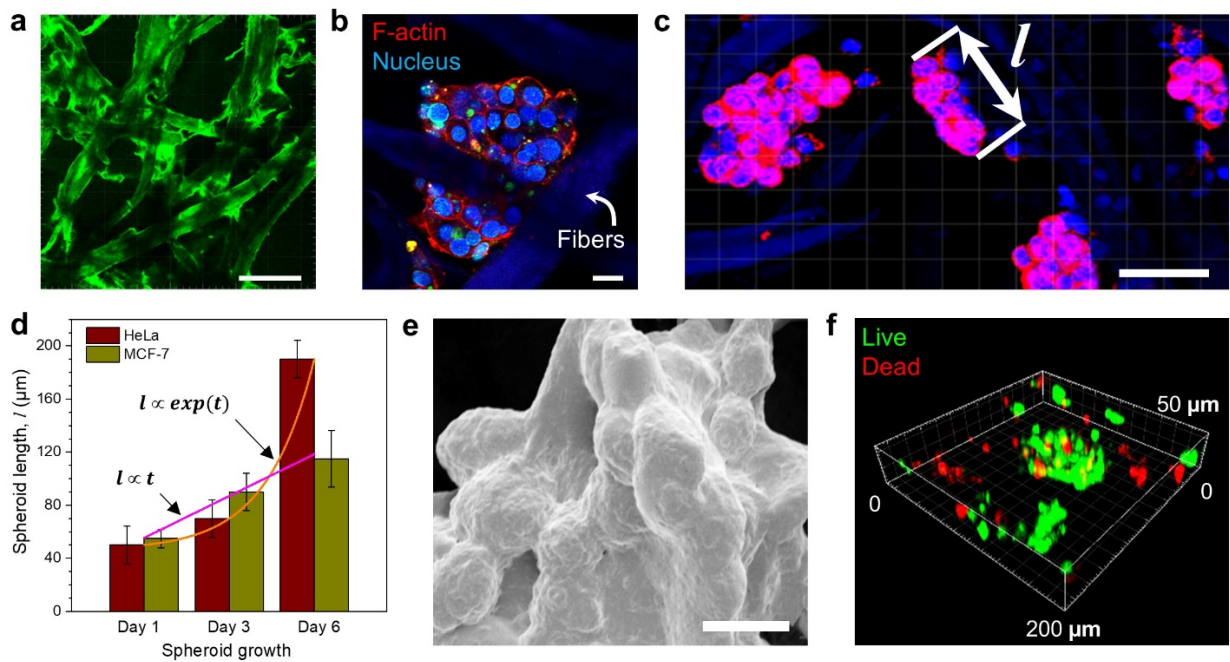

***Figure 4 | 3D spheroids within the paper platform.*** *(a) Confocal image of fibronectin-treated paper reveals fiber coating (green), pore sizes, and 3D interconnectivity of fibers. Scale bar is 100 µm. (b) Z-stack confocal image shows fairly large spheroid-like structure formed and grown within the paper pores using HeLa cells. Scale bar is 15 µm. (c & d) The length, l, of "grape-like" spheroids (Scale bar: 50 µm) was measured to investigate the increase in length of spheroids as a function of culture days for HeLa and MCF–7 cells. Values and error bars represent Mean ± S.D. (e) SEM image shows an example of accumulated cell cluster within a paper pore. Scale bar is 5 µm. (f) A 3D confocal image shows the live/dead spatial distribution of the spheroids within paper grown following thawing and cell release.*

We then conducted further investigation on the cells' post-cryopreservation 3D spread within the papers. Results showed that the cells' growth tend to fully cover the pore spaces and fibers available in their surroundings (Figure 5*e*), which corroborated our expectations that the paper-based cryopreservation fully favors the recovery of pre-cultured cells.

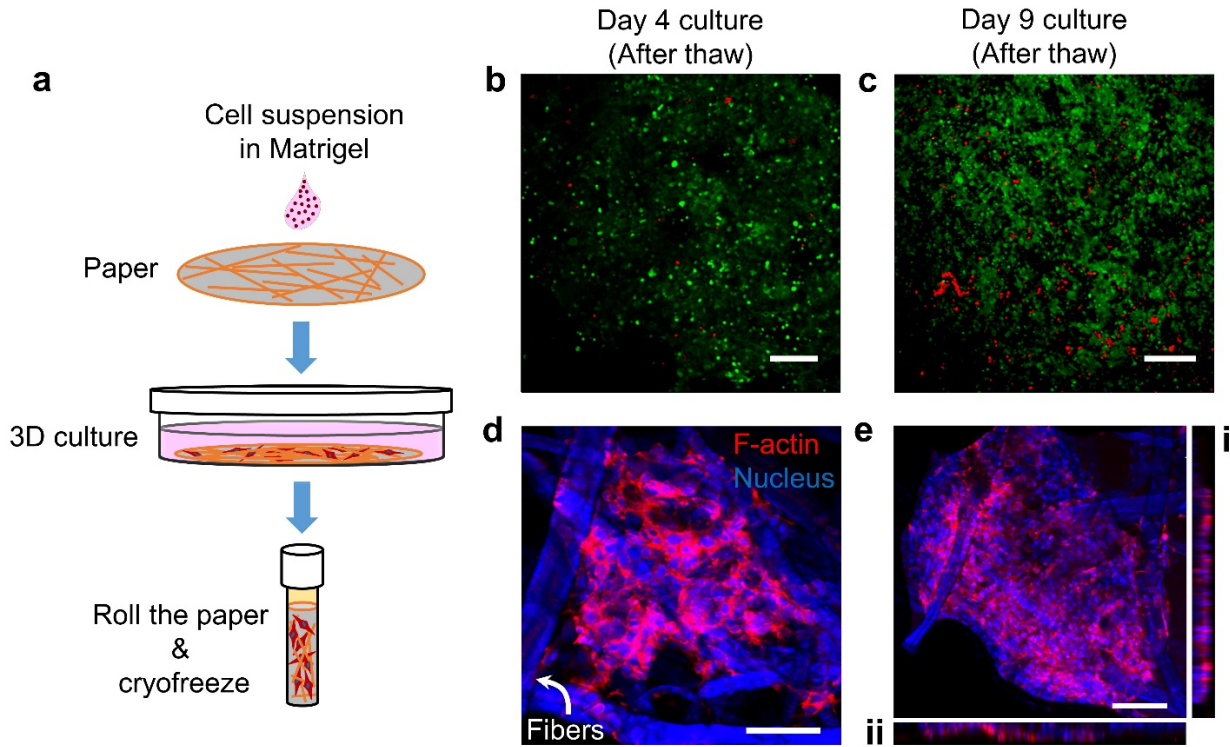

**Figure 5 | Paper-based cryopreservation of 3D cell cultures. (a)** Schematic representation of the paper-based cryopreservation method applied to the preservation of pre-cultured cells suspended in Matrigel environment. **(b & c)** Confocal images show the relative distributions of live (green) and dead (red) MCF-7 cells within paper cultured for additional 1- and 6 days after their freeze (thawed as Day 3), respectively. Scale bars are 200 μm. **(d & e)** Z-stack confocal images show the MCF-7 3D cell culture at Day 4 and Day 9, respectively. Together with the reconstituted z-section images in **(i)** and **(ii)**, it appears that their growth naturally tends to cover pore spaces and fibers available in their surroundings. Scale bar is 50 μm.

## Discussion

This is the first study to demonstrate that cells can be efficiently cryopreserved within a paper platform as an alternative to conventional cryopreservation methods. In the process of effective cell release after freeze, the significant parameter implicated in paper-based cryopreservation is chemical modification of paper fiber surfaces with fibronectin. For example, our data suggest that 10 μg/mL fibronectin concentration is sufficient to result in ~50% increment of released cells compared to untreated fibers alone. Several adhesion assays were developed and used to investigate the rates of cell-substrate bond breakages under controlled detachment forces[35, 36]. Common to these assay parameters is the initial attachment time of cells to the flat substrate, which in general defines the amount of cells released as per applied force. As such, short attachment times (<60 minutes) result in formation of fewer cell-to-surface bonds because of the round shape of the cells. Hence, when force is applied these bonds are easily dissociated resulting in detachment of cells[35, 37]. Taken together, we hypothesize that in our method fibronectin acts more like a lubricant to enhance the cell release from the paper, which is associated with their short-length (<1 minute) cell-substrate bond strengths[29, 31, 37, 38]. In addition, we assume that the highly (>50 μm) 3D porous structure of the paper[22, 39] also enhances the release of cells upon shear.

In agreement with the cell release results, paper-based cell cryopreservation showed no effect on the F-actin and tubulin integrity and organization when compared with non-frozen controls. As main components of cytoskeleton, these protein structures are known to be very sensitive to any mechanical damage of the cell membrane during cryopreservation[40]. Hence, it is important that the cryopreservation protocol does not cause any disruption in their functionality, particularly resulting from osmotic stresses[41]. Apparently, in our method the random orientation of paper fibers provides protective shield to the cells to withstand the osmotic stresses from the surrounding when they undergo the harsh freezing process. This was also evidenced by a comparative study in which the cell viability after freeze was over 2 times higher on porous fiber meshes than nonporous disks[8]. Here, we expect that the topography structure of the fibers and porosity of the paper remain intact upon freeze/thaw processes[8].

In our work, we observed that paper additionally offers—upon the release of cells after thawing—favorable environment for the growth of cell clusters within the paper which result from the remaining cells in paper and resemble 3D spheroid-like structures. Here, due to geometrical complexity of pores in heterogeneous matrices such as paper, the pores that have diameters between 50 to 200 μm shown to serve as good environment for the growth of the spheroids within the papers. In addition, we hypothesize that the interconnecting fibers serve effectively as "nests" to provide structural support for the spheroid growth deep in the paper, as revealed by scanning electron images and illustrated schematically in supplementary Figures S4*a* and S4*b*, respectively. Finally, our data showed that paper-based cryopreservation promotes the integrity and viability of cells previously cultured and grown on papers, as evidenced by their strong continuous growth within paper after thawing. Obviously, and along with previously published reports on various scaffold-based cell cryopreservations[8, 9, 27, 28], paper-based cryopreservation additionally overcomes the obstacles related to cryopreserving pre-cultured (i.e. adherent) cells on 3D porous environments; namely intracellular crystallization, dehydration injury, mechanical ruptures, and uneven cooling during the freeze.

Putting altogether, we show that the paper will utilize a unique platform for cryopreserving cells, where the fibronectin coating on the fibers and 3D porosity of the paper enhance the performance of cryopreservation in terms of high percentage of released viable cells. This makes the paper-based cryopreservation comparable to conventional cryopreservation. Additionally, the method is space saving and simple to manage, since large sheets of papers can be rolled or folded during their storage and cut into small pieces during cells' retrieval without a need to thaw the entire platform. These characteristics of the method are expected to overcome the difficulties associated with storing and managing small cryotubes of cryopreserved cells within large containers in an easy and affordable manner.

Finally, results showed that the paper-based method also promises the cryopreservation of pre-cultured cells and favors the formation and growth of spheroids-like structures from the cells remaining within the paper. These findings suggest that the method could empower the preservation of 3D cell cultures and the large-scale production of spheroids. The latter can be achieved by engineering the paper with patterned hydrophobic and hydrophilic regions so that array of "virtual" microwells are formed for preferential spheroid formations. This can be expanded even further to wider applications such as stacking the paper sheets on the top of each other to mimic different forms of *in vivo* 3D tumor models. This way, simultaneous investigation of their complex morphological and physiological characteristics in a single experiment would be possible.

## Methods

**Paper preparation.** Whatman Grade 114 resin-strengthened cellulose filter papers (Sigma-Aldrich) were used as platform in cell cryopreservation studies. After autoclaving, papers (190 µm thick) were cut into strips of about 30 mm in length and 20 mm in width. They were then either kept as untreated or submerged in a phosphate buffered saline (PBS) containing 10, 20, and 30 µg/mL concentrations of fibronectin human plasma (Sigma-Aldrich) for 20 minutes at room temperature. The excess fibronectin in the latter was removed by washing the papers twice with PBS.

**Cell loading, freezing, and thawing.** Human cervical cancer cell line (HeLa), human breast cancer cell line (MCF-7), human prostate cancer cell line (PC3), and human T-cell lymphocyte cell line (JKT) were obtained from The American Type Culture Collection (ATCC) and used in this study. They were cultured in sterile T75 tissue culture flasks (FisherScientific) using complete Dulbecco's modified essential medium (DMEM) for HeLa and MCF-7 cells and Roswell Park Memorial Institute (RPMI) medium for PC3 and JKT cells, where both media (Gibco) were supplemented with 10% fetal bovine serum (FBS; Sigma-Aldrich) and 1% penicillin-streptomycin (Pen-Strep; Sigma-Aldrich). They were then placed in a humidifying incubator at 37 ºC and 5% $CO_2$. The overall passage numbers varied between 15 and 24 for HeLa cells and between 8 and 15 for PC3, MCF-7, and JKT cells. Then, adherent HeLa, PC3, and MCF-7 cells were dissociated from the flasks when about 80% confluent using TrypLE express enzyme (Gibco) and centrifuged at 300 × g for 5 minutes, while suspended JKT cells were collected and centrifuged at 200 × g for 7 minutes. Following resuspension of cell pellets (~$10^7$ cells) in 10% FBS and 80% DMEM freezing medium complemented with 5%, for JKT cells, or 10% dimethyl sulfoxide (DMSO; Sigma-Aldrich), for HeLa, PC3, and MCF-7 cells, 300 µL of cell suspensions were pipetted onto the untreated and fibronectin-treated papers. Immediately after (<1 minute), papers were rolled and placed in standard cryotubes so that evaporation of the cell suspension is prevented. In parallel, 500 µL of cell suspensions were placed in cryotubes and used as controls. Both sets were then frozen at -80 ºC over the course of 24 hours and placed in liquid nitrogen bath (-196 ºC) for extended storage times, ranging from 3 days to 6 months. Cells loaded onto papers without cryopreservation step were used as control.

For each experiment, cryotubes were taken from liquid nitrogen and frozen cells were thawed in a 37 ºC water bath for 30 seconds. Then, the cells (suspended in freezing medium or loaded in papers) were removed from the cryotubes and resuspended/placed in centrifuge tubes containing 10 mL of warm DMEM medium. To release the cells from paper, the tubes were shaken gently for about 20 seconds. Then, the cell suspensions were centrifuged and the cell pellets were resuspended in a fresh complete culture medium. Cell counting chamber slides (Invitrogen) and 96-well culture plates (FisherScientific) were used for viability and proliferation assays, respectively, as explained in details below. Unless otherwise stated, all assays were repeated twice and performed using triplicate samples.

**Cell viability and proliferation.** After freezing and thawing, live/dead fluorescent cell viability imaging assay (Invitrogen) was used to visually assess (as control) the distribution of live/dead cells (HeLa) within the papers—prior to and after their release—using confocal microscopy. Meanwhile, in parallel released cells from papers were removed from DMEM medium and washed three times with PBS. Following their resuspension in fresh culture medium, trypan blue exclusion assay (Sigma-Aldrich) was applied to directly count the number of released live and dead cells using Countess II FL (FisherScientific) automated cell counter and their averaged proportions were converted to percentages and recorded as bar charts for comparison.

WST-1 test (Sigma-Aldrich) was used to quantify the proliferation of released cells. Here, cells were seeded in 96-well culture plates at concentrations of ~$10^4$ cells/mL per well for 3 days and WST-1 was performed at days 1, 2, and 3. Following their incubations at 37 °C and 5% $CO_2$, at each day, Varioskan Flash plate reader (FisherScientific) was used to determine the cell growth by measuring the changes in the optical densities at 440 nm (OD440) in correlation with the confluence of proliferated cells. In parallel, the background was measured at OD640 and subtracted from each OD440 measurement. Then, the mean OD values were compared to the proliferation of cells cryopreserved using conventional method, with and without fibronectin as additive to cryopreservation medium, and recorded as bar charts.

Following the proliferation measurements, cells were fixed with paraformaldehyde (Sigma-Aldrich) in PBS (2%, v/v) for 10 minutes at room temperature and permeabilized with Triton X-100 (Sigma-Aldrich) in PBS (0.5%, v/v) for 15 minutes. Then, bovine serum albumin (BSA; Sigma-Aldrich) in PBS (1%, v/v) was used for 20 minutes to block any non-specific staining. After blocking, F-actin, tubulin, and nuclei staining was done using rhodamine phalloidin conjugated to red-orange fluorescent dye (Cytoskeleton), anti-Alpha tubulin antibodies conjugated to green dye (Abcam), and Hoechst 33342 conjugated to blue dye (Invitrogen), respectively, for 20 minutes at room temperature in dark. Finally, their images were taken using confocal microscopy. As a result, the morphological changes in their structures were visually examined and compared with the proliferation response of conventionally cryopreserved cells.

**Spheroid formation and size evaluation.** After thaw/release process, the papers were re-submerged in a complete culture medium (DMEM supplemented with 10% FBS and 1% Pen-Strep) and incubated at 37 °C and 5% $CO_2$ for 3 days. After every other day, live/dead assays were performed to visually verify the viability of the spheroids as control. In parallel, after washing papers gently with PBS, cells were stained for F-actin, tubulin, and nuclei for 20 minutes at room temperature in dark following the procedure explained in previous section. Then, the papers were mounted on microscopy slides, with their top or bottom sides facing down, using a drop of mounting medium (Vectashield) and left at 4 °C for overnight. Finally, the cells within the papers were imaged using confocal and scanning electron microscopy, as explained in detail in the imaging section.

The size evaluation of spheroids was carried out by measuring their apparent length in the projection of z-stack images. HeLa and MCF-7 cells were used as model. With 2-3 spheroids per measurement, the spheroid lengths following their growth at days 1, 3, and 6 were measured and their mean values were recorded as bar charts.

**Pre-cultured cell preservation.** HeLa and MCF-7 cells (~$10^7$ cells/mL) were added to 100% Matrigel (Sigma-Aldrich) at 4 °C (placed in ice) and subsequently their cell suspensions in Matrigel were pipetted onto the papers[22]. Then, papers were left for 10 minutes at room temperature for the Matrigel to solidify. Following, papers were submerged in a complete culture medium for 3 days at 37 ºC and 5% $CO_2$ to allow the proliferation of cells. Then, papers were washed three times in PBS, rolled and placed in standard cryotubes containing DMEM freezing medium and frozen at -80 ºC for overnight. After their preservation at -196 ºC for 3 days, cells were thawed, and papers were removed from the cryotubes and resuspended in complete culture medium for additional 8-day cell proliferation. Then, cells were fixed and stained for F-actin and nucleus following the methodology explained before. Finally, they were imaged using confocal microscopy.

**Scanning electron and confocal microscopy imaging.** Cambridge S360 scanning electron microscope (Leica) was used to image the cells and beads (FisherScientific) within the papers at 5 kV accelerating voltage. The images were acquired digitally using UltraScan software (UltraScan Project). Prior to the imaging, cells were fixed in paraformaldehyde in PBS (2%, v/v) for 10 minutes, dehydrated using increasing concentrations (from 20 to 100%) of ethanol solutions and left air dry overnight at room temperature. Imaging was carried with coating samples with conductive layer.

Confocal microscopy imaging of cells was carried with FV1000 inverted confocal laser scanning microscope (Olympus) using blue (405 nm), green (488 nm), and red (612 nm) excitation wavelengths. 10× air objective lens was used to image up to 160 μm deep in paper. The z-stack imaging was performed in 5 μm increments and their 3D projections were created using Imaris software (Bitplane).

**Statistical analysis.** Statistical analysis was performed with Origin software (OriginLab) using two-way analysis of variance (ANOVA) to evaluate differences between viability of released cells after their cryopreservation using paper and conventional methods. A P-value of <0.05 was considered statistically significant.

## Acknowledgements

This study was financially supported by NYU Abu Dhabi (NYUAD) and in part by Sheikh Hamdan Award for Medical Science Research Grant (Dubai, UAE) under cycle 2017-2018. We thank Dr. Rachid Rezgui for the technical support and Mrs. Christina Johnson for creating the schematic in Figure 1a. We acknowledge Core Technology Platforms at NYUAD for use of its microscopy facilities.

## Author contributions

R.A. and M.A.Q. designed the experiments, R.A. and P.S. performed the cell cryopreservation experiments and the confocal microscopy imaging, M.D. performed the SEM imaging, R.A., P.S., M.D., and M.A.Q. analyzed the data, R.A., P.S., M.D., and M.A.Q. prepared the figures, R.A., M.D., and M.A.Q. wrote the paper.

## Additional Information

**Supporting information** Four figures (S1-S4) are presented as supporting material for this paper.

**Competing interests** The authors of this paper declare no competing interests.

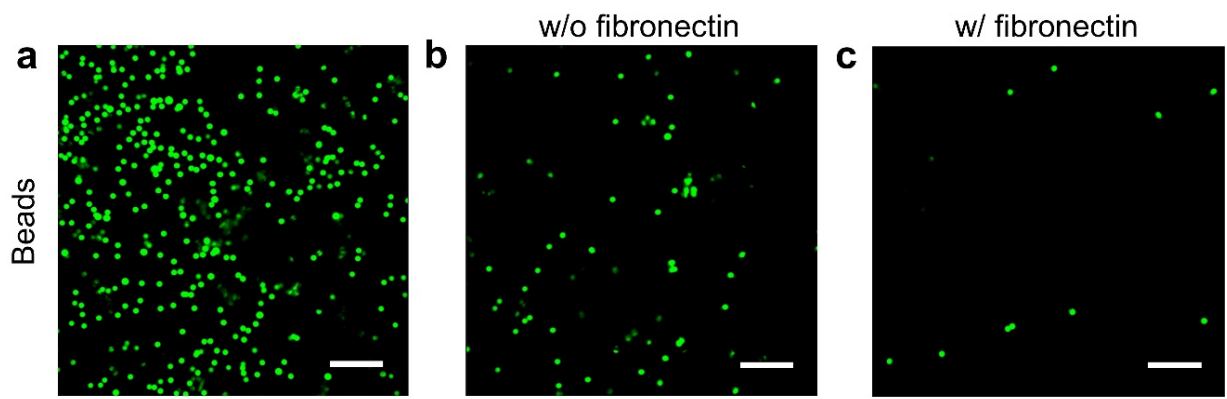

**Supplementary Figure S1 | Characterization of bead release from paper. (a-c)** Confocal images revealed that papers treated with 10 µg/mL concentration of fibronectin favor more release of beads as compared to untreated ones. Scale bars are 200 µm.

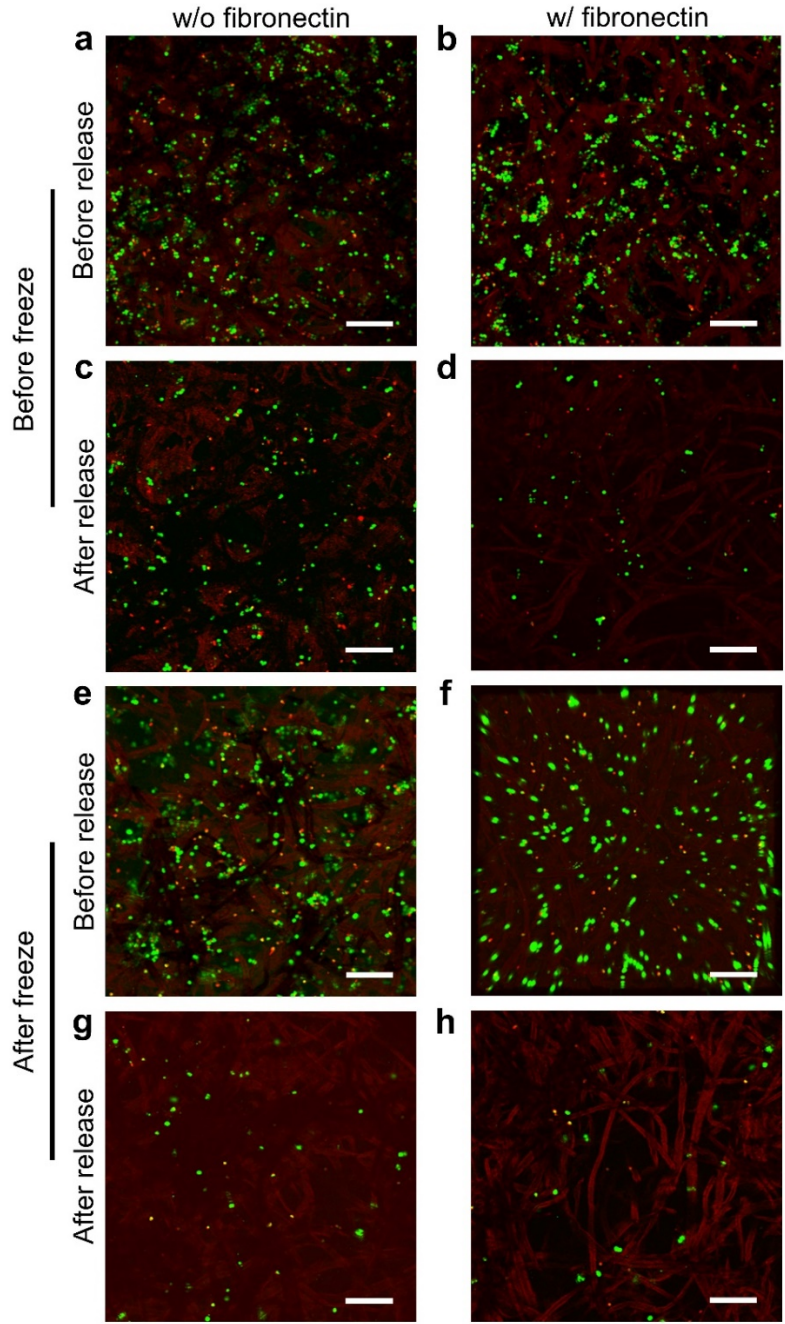

**Supplementary Figure S2 | Characterization of cell release from paper. (a-h)** Following their load onto untreated and 10 µg/mL fibronectin-treated papers, the relative distribution of live (green) and dead (red) HeLa cells within papers visually showed no difference for **(a & b)** before and **(e & f)** after their freeze, respectively. However, investigation of the role of fibronectin in the release of live/dead cells within the papers **(c & d)** prior to and **(g & h)** after their freeze confirmed that coating paper fibers with fibronectin enhances the effective release of more viable cells compared to untreated ones. As for the release of dead cells, fibronectin had no apparent effect compared to the non-frozen controls. We hypothesize that in both scenarios (i.e. w/ and w/o fibronectin) the dead cells were more likely entrapped within the 3D fiber network of the paper so that they washed off more easily. Scale bars are 200 µm.

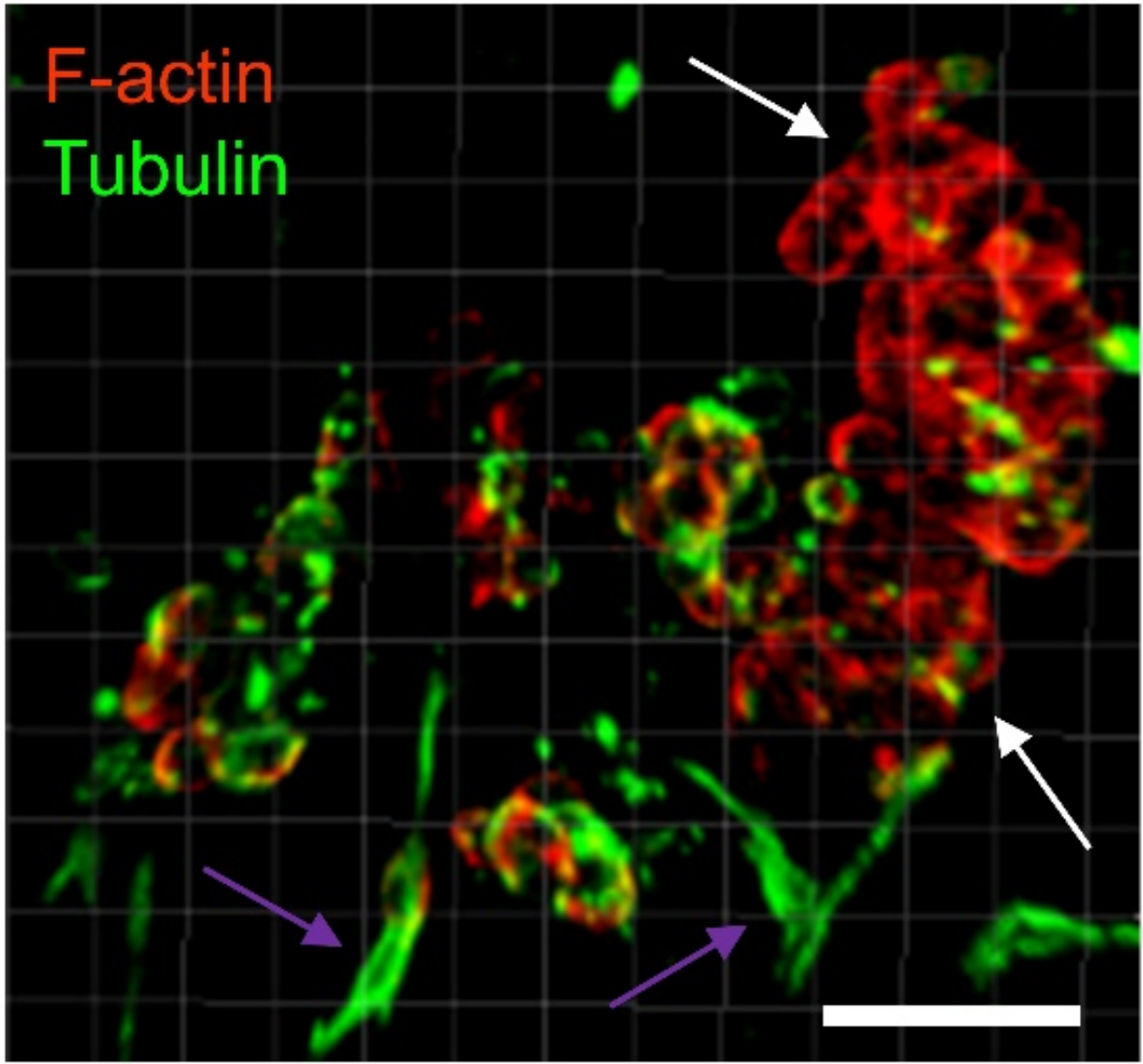

**Supplementary Figure S3 | Multiple 3D cell formations within paper.** Z-stack confocal image reveals the formation of 3D cell cultures (purple arrows) and spheroids (white arrows) of cryopreserved HeLa cells after their culture for 1 day. The tubulin appeared to be more extended in 3D cell cultures, whereas the F-actin was more condensed in spheroids. These formations were observed when low concentrations (<$10^7$ cells/mL) of cells were loaded onto papers prior to cryofreeze. Scale bar is 80 μm.

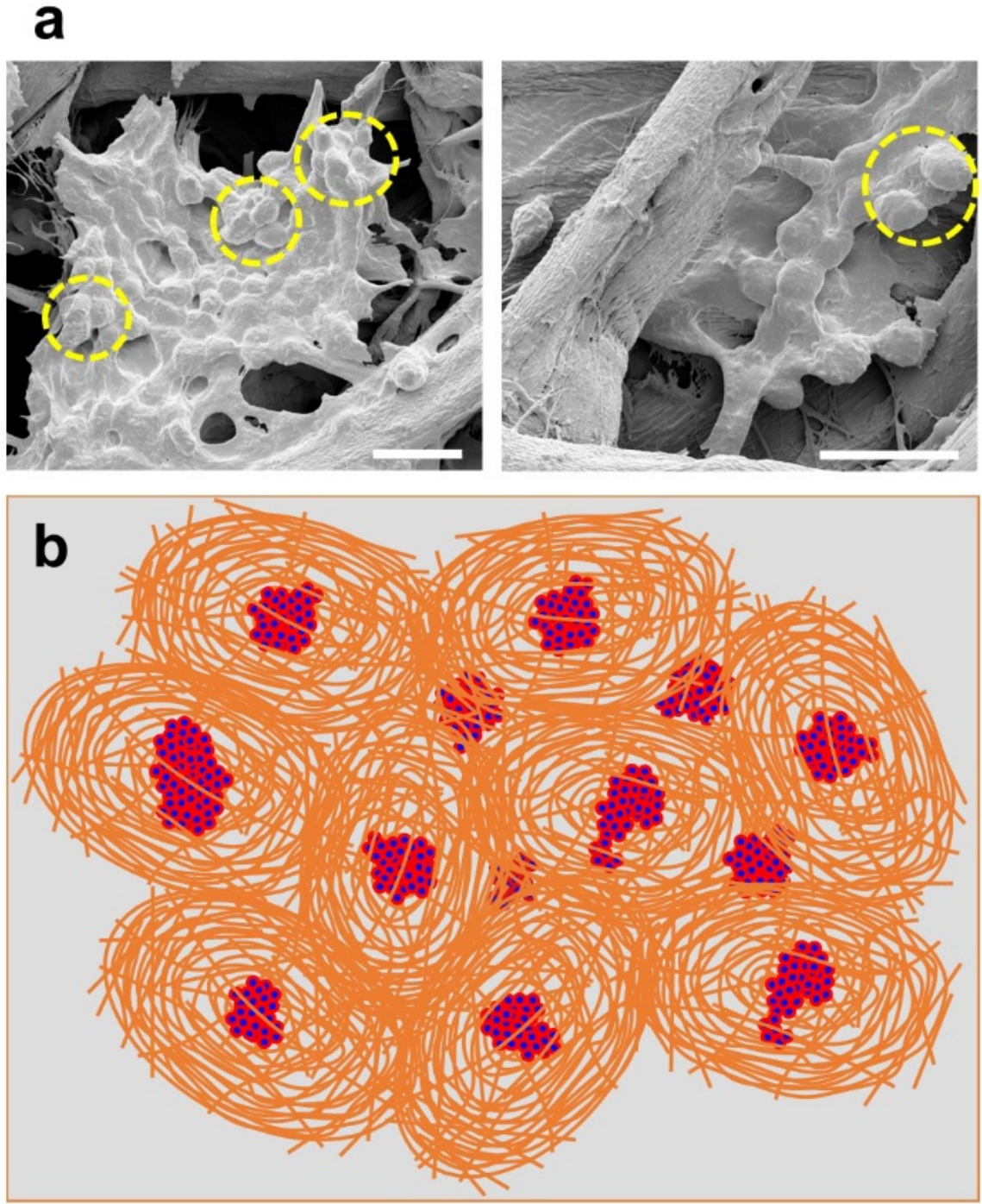

**Supplementary Figure S4 | Interconnecting fibers in conjunction to the growth of spheroids.**
**(a)** SEM images reveal the existence of post-cryopreserved cells within the paper pores, wherein they seemingly act as connecting bridges between the paper fibers. Example cell clusters (dashed yellow circles) are clearly distinguishable among them. Scale bars are 20 μm. **(b)** Schematic representation of the interconnecting fibers (orange) within paper that supposedly serve as nests to support the growth of 3D spheroids (blue = nucleus and red = F-actin) within a paper following their cryopreservation.